\documentclass[twocolumn,superscriptaddress,tightenlines,tightenlines, prl]{revtex4-1}
\usepackage{amsthm}
\usepackage{amsmath}
\usepackage{graphicx}
\usepackage[usenames,dvipsnames]{color}
\usepackage[colorlinks=true,citecolor=magenta,urlcolor=blue]{hyperref}
\usepackage{colortbl}
\usepackage{color}
\usepackage{multirow}
\usepackage{hhline}
\usepackage{epstopdf}

\newcolumntype{C}[1]{ >{ \centering\arraybackslash}p{#1}}

\newcommand{\X}{$\mathsf{X}$}
\newcommand{\Z}{$\mathsf{Z}$}

\newcommand{\sz}[1]{s_{\mathsf{Z}, #1}}

\newcommand{\phiz}{\phi_{\mathsf{Z}}}

\newcommand{\be}{\begin{equation}}
\newcommand{\ee}{\end{equation}}

\newcommand{\del}[1]{}

\newcommand{\sket}[1]{{\ensuremath{\lvert#1\rangle}}}
\newcommand{\lket}[1]{{\ensuremath{\left\lvert#1\right\rangle}}}
\newcommand{\ket}[1]{\if@display\lket{#1}\else\sket{#1}\fi}

\newcommand{\sbra}[1]{{\ensuremath{\langle#1\rvert}}}
\newcommand{\lbra}[1]{{\ensuremath{\left\langle#1\right\rvert}}}
\newcommand{\bra}[1]{\if@display\lbra{#1}\else\sbra{#1}\fi}

\newcommand{\sbraket}[2]{{\ensuremath{\langle#1\rvert#2\rangle}}}
\newcommand{\lbraket}[2]{{\ensuremath{\left\langle#1\!\left\rvert\vphantom{#1}#2\right.\!\right\rangle}}}
\newcommand{\braket}[2]{\if@display\lbraket{#1}{#2}\else\sbraket{#1}{#2}\fi}

\newcommand{\sketbra}[2]{{\ensuremath{\lvert #1\rangle\!\langle #2\rvert}}}
\newcommand{\lketbra}[2]{{\ensuremath{\left\lvert #1\right\rangle\!\!\left\langle #2\right\rvert}}}
\newcommand{\ketbra}[2]{\if@display\lketbra{#1}{#2}\else\sketbra{#1}{#2}\fi}

\begin{document}
\title{Secure quantum key distribution over 421 km of optical fiber}

\author{Alberto Boaron} \email{alberto.boaron@unige.ch}
\author{Gianluca Boso}
\author{Davide Rusca}
\author{C\'edric Vulliez}
\author{Claire Autebert}
\author{Misael Caloz}
\author{Matthieu Perrenoud}
\affiliation{Group of Applied Physics, University of Geneva, Chemin de Pinchat 22, CH-1211 Geneva 4, Switzerland}
\author{Ga\"etan Gras}
\affiliation{Group of Applied Physics, University of Geneva, Chemin de Pinchat 22, CH-1211 Geneva 4, Switzerland}
\affiliation{ID Quantique SA, 3 Ch. de la Marbrerie, CH-1227 Carouge, Switzerland}
\author{F\'elix Bussi\`eres}
\affiliation{Group of Applied Physics, University of Geneva, Chemin de Pinchat 22, CH-1211 Geneva 4, Switzerland}

\author{Ming-Jun Li}
\affiliation{Corning Incorporated, Corning, NY 14831, United States}
\author{Daniel Nolan}
\affiliation{Corning Incorporated, Corning, NY 14831, United States}
\author{Anthony Martin}
\affiliation{Group of Applied Physics, University of Geneva, Chemin de Pinchat 22, CH-1211 Geneva 4, Switzerland}
\author{Hugo Zbinden}
\affiliation{Group of Applied Physics, University of Geneva, Chemin de Pinchat 22, CH-1211 Geneva 4, Switzerland}

\begin{abstract}
We present a quantum key distribution system with a 2.5~GHz repetition rate using a three-state time-bin protocol combined with a one-decoy approach.
Taking advantage of superconducting single-photon detectors optimized for quantum key distribution and ultra low-loss fiber, we can distribute secret keys at a maximum distance of 421~km and obtain secret key rates of 6.5~bps over 405~km.
\end{abstract}

\maketitle

The first experimental demonstration of quantum key distribution (QKD) was over a short distance of 32~cm on an optical table~\cite{Bennett1992}.
Since then, there has been a continuous progress on the theoretical and technological side such that nowadays commercial fiber-based systems are available~\cite{IDQ} and the maximum distance has been pushed up to 400~km with academic systems~\cite{Yin2016}.
Recently, the feasibility of satellite-based QKD has been demonstrated~\cite{Liao2017}, opening the door for world-wide key distribution for the lucky owners of satellites~\cite{Liao2018}.

The maximum distance of fiber-based systems is mainly limited by two factors.
On one hand, the detector noise which, due to the exponential decrease of the signal, eventually becomes the dominant source of error and abruptly ends the possibility to extract a key.
On the other hand, in the limit of arbitrarily low detector noise, it is the maximal acceptable key accumulation time.
Indeed, taking into account finite key analysis, a secret key cannot be extracted with high confidence for short blocks of raw key.
A system with high pulse rate and efficient detectors can therefore push this limit a bit further.

In this paper, we present an experiment that takes advantage of state-of-the-art performance on all fronts to push the limits to new heights.
We rely on a new 2.5~GHz clocked setup~\cite{Boaron2018}, low-loss fibers, in-house-made highly efficient superconducting detectors~\cite{Caloz2018} and last but not least a very efficient one-decoy state scheme~\cite{Rusca2018a}.
Finally, we achieve an improvement of the secret key rate (SKR) by four orders of magnitudes with respect to a comparable experiment over 400~km.

We implement the protocol presented in Boaron \textit{et al.}~\cite{Boaron2018}.
For the sake of simplicity of the setup, we use a three-state time-bin scheme: two states in the \Z{} basis (a weak coherent pulse in the first or the second time-bin, respectively) and one state in the \X{} basis (a superposition of two pluses in both time-bins).
Moreover, we employ only two detectors.
The finite-key security analysis of this scheme is briefly outlined below and detailed in Rusca \textit{et al.}~\cite{Rusca2018b}.
In order to be robust against photon number splitting attacks over long links (with high total loss) the decoy state method~\cite{Wang2005, Lo2005} is applied.
In particular we use the one-decoy state approach, which was shown to be optimal for block sizes smaller than $10^8$~bits~\cite{Rusca2018a}.
All pulses have random relative phase in order to render coherent attacks inefficient.


\begin{figure*}
\begin{minipage}[c]{0.72\textwidth}
\includegraphics[width = 0.9\columnwidth]{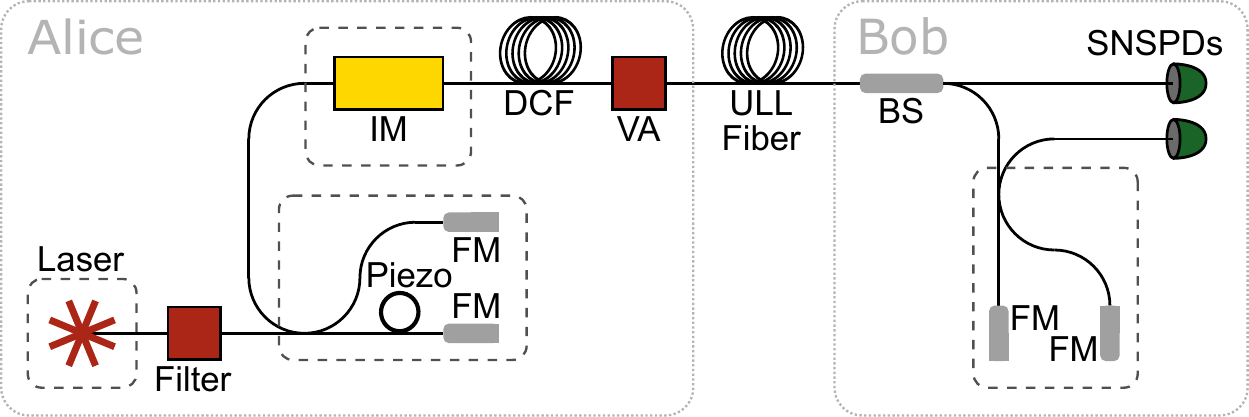}
\end{minipage}\hfil
\begin{minipage}[c]{0.28\textwidth}
\caption{Schematics of the experimental setup. Laser: 1550~nm distributed feedback laser; Filter: 270~pm bandpass filter; Piezo: piezoelectric fiber stretcher; FM: Faraday mirror; IM: intensity modulator; DCF: dispersion compensating fiber; VA: variable attenuator; ULL fiber: ultra low-loss single-mode fiber; BS: beamsplitter; SNSPDs: superconducting nanowire single-photon detectors. Dashed lines represent temperature stabilized boxes.}
\label{fig.implementation}
\end{minipage}
\end{figure*}

Figure~\ref{fig.implementation} schematically shows our experimental realization.
Alice's and Bob's setups are situated in two separated laboratories 20~m apart.
Each of them is controlled by a field programmable gate array (FPGA).

Alice uses a phase-randomized diode laser pulsed at 2.5~GHz.
Phase randomness is achieved by switching the current completely off between the pulses~\cite{Kobayashi2014}.
The pulses then pass through an unbalanced Michelson interferometer (200~ps delay).
One of its arms is equipped with a piezoelectric fiber stretcher to adjust the phase.
The different qubit states are now encoded by a lithium niobate intensity modulator controlled by the FPGA.
The qubit states and the pulse energies (signal or decoy state) are chosen at random.
For this purpose, we rely on a quantum random number generator [ID Quantique, Quantis] which supplies 4~Mbps of random bits which are expanded to 40~Gbps using the NIST SP800-90 recommended AES-CTR cryptographically secure pseudo-random number generator.

Bob's choice of measurement basis is made passively by a beamsplitter.
In the \Z{} basis, the photons are directly sent to a single-photon detector that measures their arrival time.
This basis is used to generate the raw key.
In the \X{} basis, used to estimate the eavesdropper information, an unbalanced interferometer identical to that of Alice allows to measure the coherence between two consecutive pulses.
Only one detector is employed at the output of the interferometer.

The quantum channel (QC) is composed of spools of SMF-28$^{\mbox{\scriptsize{\textregistered}}}$ ultra low-loss (ULL) single-mode fiber (SMF) [Corning] which has an attenuation of about 0.16~dB/km (0.17~dB/km including the connections loss) and a positive chromatic dispersion of around 17~ps\,nm$^{-1}$\,km$^{-1}$.
The ULL fiber consists of a pure silica core and a fluorine doped cladding.
To reduce the impact of the chromatic dispersion, we pre-compensate it with dispersion compensation fiber (DCF) fabricated by Corning Inc. placed on Alice's side.
The DCF dispersion is around -140~ps\,nm$^{-1}$\,km$^{-1}$ and its attenuation is about 0.5~dB/km.

The synchronization and communication between Alice's and Bob's devices is performed through a communication link, denoted as service channel (SC), based on small form-factor pluggable (SFP) transceivers connected through a short 50~m duplex fiber.
For practicality, we use this fiber for all QC lengths.
However, a SC of the same length as the QC (implemented with optical amplifiers) would offer better stability.
Anyway, we compensate actively the fluctuations of the path length difference between the QC and the SC.
For this purpose, the detectors signals are sampled at 10~GHz (i.e. only half of the bins are used for the sifting).
The temporal tracking is performed by minimizing the ratio between the detections in the inactive and active bins.
At the distances under study, we observed drifts having a sinusoidal behavior over one day, with amplitudes up to about 10~ns (which correspond to a 0.5~K difference in the average fiber temperature at 400~km).
The intrinsic phase stability of our interferometers exceeds 10 minutes.
Still, an automatic feedback loop also stabilizes the relative phase between Alice's and Bob's interferometers using the quantum bit error rate (QBER) in the \X{} basis as an error signal.
The temporal tracking and the phase stabilization work in real time for distances up to 400~km.
However, at the maximal distance (421~km), given the low detection rate, the statistical fluctuations of the error signal become too important to stabilize in real time.
Therefore, we interrupt data acquisition after each block of error correction (EC) (about half an hour of acquisition) in order to perform an adjustment with a higher power of Alice's signal.

The detection is done with two custom-made molybdenum silicide superconducting nanowire single-photon detectors (SNSPDs) cooled at 0.8~K~\cite{Caloz2018}.
For SNSPDs, reducing the noise of the detectors implies filtering out black-body radiation present in the optical fiber leading to the detector.
The black-body radiation around the laser wavelength (1550.92~nm) is eliminated using a standard 200~GHz fibered dense wavelength division multiplexer bandpass filter cooled to 40~K.
Infrared light above 1550~nm is filtered by coiling the optical fiber just before the detector~\cite{Smirnov2015}.
In this way, we achieve a dark count rate (DCR) of 0.1~Hz, which is close to the intrinsic DCR of the detectors.
The maximum efficiencies of our detectors are between 40 to 60\%, depending on the detector and on the filtering configuration.
Because of the meander structure of the SNSPDs, the detection efficiency depends on the input polarization (the ratio between the minimum and maximum efficiencies is about $1/2$).
This leads to slow variations of the detection rate, since we adjust the polarization of the light at the beginning of the runs, but do not perform any further adjustment during the acquisition.
The system timing jitter of the detectors is lower than 40~ps.


The model of our protocol consists in a modification from the already proven to be secure three-state protocol~\cite{Fung2006, Tamaki2014, Mizutani2015}.
The difference stands in the fact that we have only one detector in the \X{} basis.
Therefore, we do not have access to all measurements outcomes of the standard protocol.
However, this does not affect the security of the protocol as demonstrated in Rusca \textit{et al.}~\cite{Rusca2018b}.
Note that the proof covers the security against collective attacks.
However, given the phase-randomization of the states sent by Alice, the results can be extended to coherent attacks using techniques such as Azuma's inequality~\cite{Azuma1967, Boileau2005, Tamaki2009} or De Finetti's theorem~\cite{Caves2002, Konig2005}.

The secure key bits per privacy amplification block is given by~\cite{Rusca2018a}:
\begin{align}\label{eq:skl}
l \leq & \sz{0} + \sz{1}(1-h(\phiz)) - \lambda_\text{EC} \nonumber\\
& - 6\log_2(19/\epsilon_\text{sec}) - \log_2(2/\epsilon_\text{cor}),
\end{align}
where $\sz{0}$ and $\sz{1}$ are the lower bound on the number of vacuum and single-photon detections in the \Z{} basis, $\phiz$ is the upper bound on the phase error rate, $\lambda_\text{EC}$ is the total number of bits revealed during the EC, and $\epsilon_\text{sec} = 10^{-9}$ and $\epsilon_\text{cor} = 10^{-9}$ are the secrecy and correctness parameters, respectively.

\begin{table*}
\begin{tabular}{C{1.4cm}C{1.4cm}|C{1.4cm}C{1.4cm}|C{1.8cm}C{1.8cm}|C{1.4cm}C{1.4cm}C{1.8cm}C{1.8cm}}
\hline \hline
length&attn&$\mu_1$&$\mu_2$&block size&block time&QBER \Z{}&$\phiz$&RKR&SKR\\
(km)&(dB)&&&&(h)&(\%)&(\%)&(bps)&(bps)\\
\hline
251.7 & 42.7 & 0.49 & 0.18 & $8.2 \cdot 10^{6}$ & 0.20 & 0.5 & 2.2 & $12 \cdot 10^{3}$ & $4.9 \cdot 10^{3}$ \\
302.1 & 51.3 & 0.48 & 0.18 & $8.2 \cdot 10^{6}$ & 1.17 & 0.4 & 3.7 & $1.9 \cdot 10^{3}$ & $0.79 \cdot 10^{3}$ \\
354.5 & 60.6 & 0.35 & 0.15 & $6.2 \cdot 10^{6}$ & 14.8 & 0.7 & 1.8 & $117$ & $62$ \\
404.9 & 69.3 & 0.35 & 0.15 & $4.1 \cdot 10^{5}$ & 6.67 & 1.0 & 4.3 & $17$ & $6.5$ \\
421.1 & 71.9 & 0.30 & 0.13 & $2.0 \cdot 10^{5}$ & 24.2 (12.7*) & 2.1 & 12.8 & $2.3$ ($4.5$*) & $0.25$ ($0.49$*) \\
\hline \hline
\end{tabular}
\caption{\label{tab.results} Overview of experimental parameters and performance for different fiber lengths. *Data considering only the duration of the data transmission.}
\end{table*}


We performed key exchanges with fiber lengths between 252 and 421~km.
For every distance we optimized the following experimental parameters to maximize the SKR.
On Alice's side, we varied the probability of choosing the \Z{} and \X{} basis, the mean photon number of the two decoy states $\mu_1$ and $\mu_2$ and their respective probabilities.
On Bob's side, we used different detectors following a trade-off between high efficiency and low DCR.
The latter criterion becomes increasingly important with increasing distances.
For simplicity, Bob's probability of choosing the \Z{} and \X{} basis was kept constant to $1/2$, which is a good value at long distances to minimize the penalty due to the finite-key analysis in both bases.

\begin{figure}[b]
\includegraphics[width = 0.9\columnwidth]{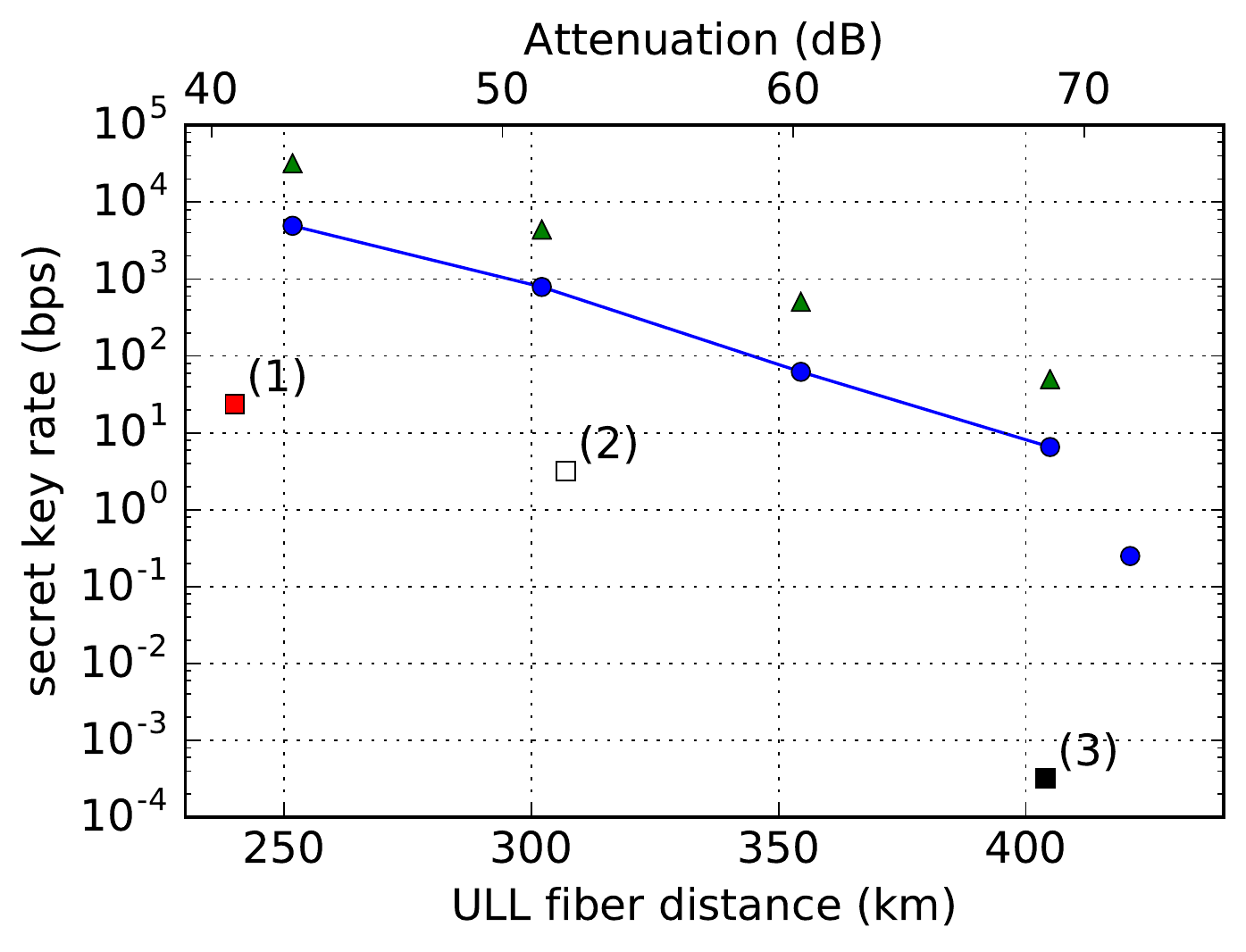}
\caption{
Circles: experimental final SKR versus distance.
Triangles: simulation of an idealized BB84 protocol with the same block sizes as the corresponding experimental points.
Squares: results of other long-distance QKD experiments using ULL fibers: (1) BB84, B. Fr\"olich \textit{et al.}~\cite{Frohlich2017}; (2) Coherent one-way, B. Korzh \textit{et al.}~\cite{Korzh2015}; (3) Measurement-device-independent QKD, H.-L. Yin \textit{et al.}~\cite{Yin2016}. The upper axis is obtained by considering an attenuation of 0.17~dB/km.}
\label{fig.SKRvsDST}
\end{figure}

\begin{figure}[b]
\includegraphics[width = 0.9\columnwidth]{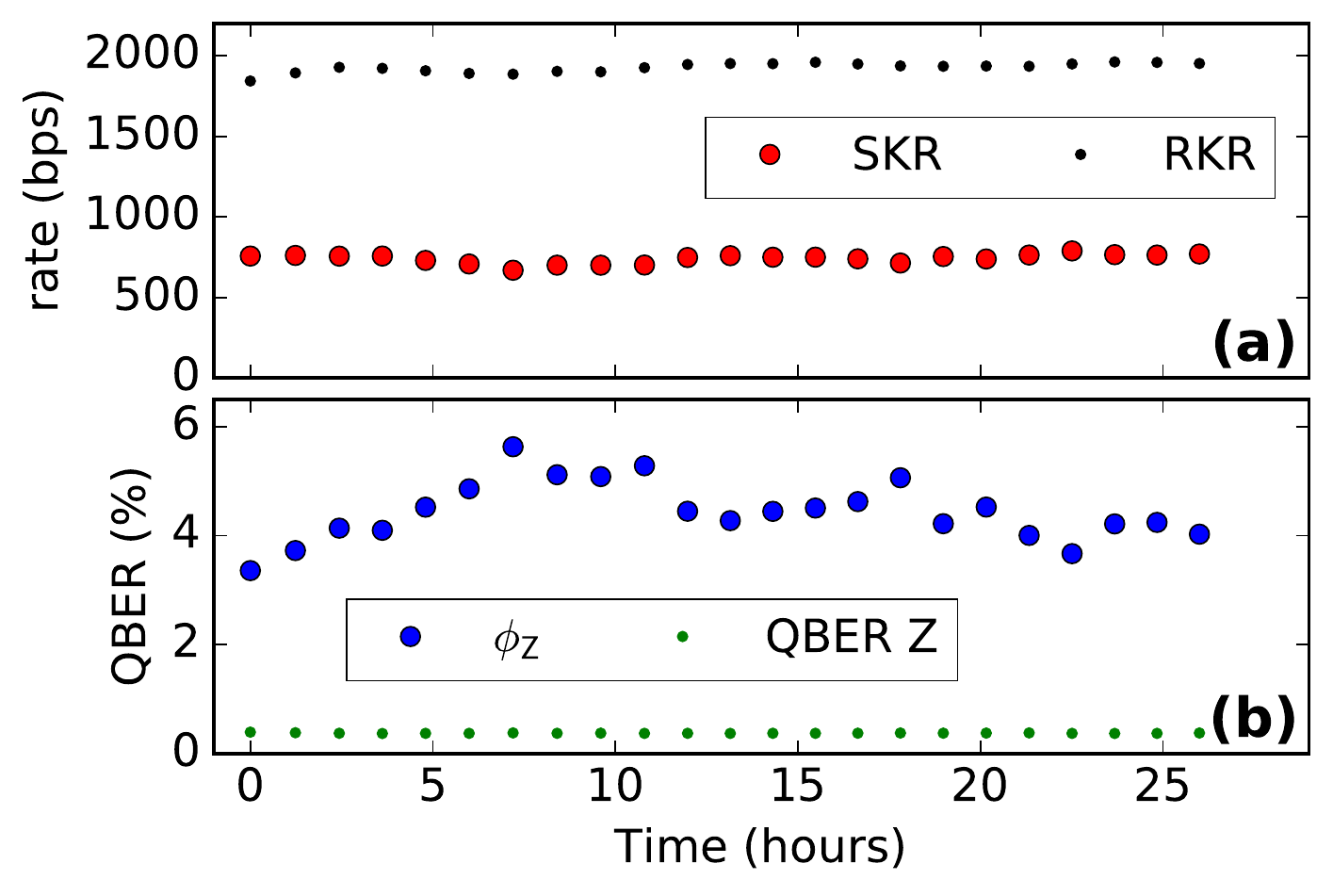}
\caption{System stability over more than 24~h for a distance of 302~km of ULL SMF. (a) RKR, SKR, and (b) corresponding QBER in the \Z{} basis and $\phiz$ as a function of time.}
\label{fig.long_measurement}
\end{figure}

Table~\ref{tab.results} summarizes the experimental settings and the results obtained for each distance.
Figure~\ref{fig.SKRvsDST} shows the SKR as a function of the distance.
At shorter distances, the QBER is mainly due to the imperfect preparation of the states by Alice (in particular due to limited extinction ratio of the intensity modulator).
Indeed, the errors caused by the timing jitter of the detectors should not exceed 0.1\% thanks to the small and Gaussian-shaped timing jitter of SNSPDs.
Given our detection method with a 10~GHz sampling (the bins are 100~ps wide), a detection has to occur 150~ps away from the central timing to generate an error.
For a 40~ps jitter, this corresponds to more than 3$\sigma$, leading to an error probability smaller than 0.1\%.
(We would expect this value to be at least one order of magnitude bigger for avalanche photodiode single-photon detectors~\cite{Boaron2018}.)

The contribution of the DCR to the QBER becomes significant only above 350~km.
At this distance the imperfect temporal tracking due to faster variation and lower error signal starts to contribute as well.
Similarly, the phase error rate is additionally affected by the imperfect stabilization of the interferometers.

For 405~km and 421~km, in order to keep the acquisition time shorter than one day, we reduced the privacy amplification block size by more than a factor of ten compared to shorter distances.
The finite-key analysis leads therefore to lower SKRs that are about half of the SKRs one would obtain in the case of infinite keys.

To obtain the 421~km point, we run the system over three periods corresponding to a total of 24.2~h of acquisition time, including the necessary interruptions for alignment.
39 EC blocks were generated of which we kept the 25 blocks with the best performance.
This allowed us to extract 22124 secret bits, which corresponds to a SKR of 0.25~bps.
Considering only the time necessary to exchange the 25 EC blocks (12.7~h), we obtain a SKR of 0.49~bps.

To demonstrate the long-term operation capability of our system, we run it over a continuous period of more than 24~h at a transmission distance of 302~km.
The phase stabilization and temporal alignment were performed automatically by the control software.
The relevant experimental results are shown in figure~\ref{fig.long_measurement} as a function of time.
Fluctuations of the raw key rate (RKR) are mainly due to polarization fluctuations of the signal arriving at Bob's side.

Figure~\ref{fig.SKRvsDST} also shows a comparison of our experimental results with other QKD realizations.
421~km is the maximal transmission distance reported for a QKD system in fiber.
Compared to the previous record~\cite{Yin2016}, at 405~km, the rate is improved by four orders of magnitude.
Moreover, our acquisition times, shorter than a day, are still of practical utility.

In order to appreciate the performance of our system with respect to a perfect one, we simulated (for the same distances and block sizes as our experimental points) the SKRs of an idealized BB84 system with no DCR, 0\% of QBER and 100\% detection efficiency (represented as triangles on figure~\ref{fig.SKRvsDST}).
Most of the difference is due to the lower detection efficiency in our experiment.
Indeed, if we took it into account, the simulated and experimental points would almost overlap.
Therefore, we can conclude that our simplifications of the protocol (three-state) and the implementation (with only one detector in the \X{} basis) do not significantly affect the performance. Except for the detection efficiency, our system is close to an ideal system.

How far could one still increase the transmission distance of QKD?
With an ideal, noiseless implementation, the limiting factor is in the end the minimum block size needed to still extract a secret key with good confidence.
Given that the number of detected photons decreases exponentially with distance, the resulting, necessary exponential increase of the accumulation time cannot be satisfactorily mitigated by an increased pulse repetition rate.
We simulate a system with the following properties: BB84 protocol, 10~GHz repetition rate, 100\% detector efficiency, 0~Hz DCR and $\epsilon_{\rm{sec}} = 10^{-9}$.
For this system, a constraint of 1~day of acquisition leads to a maximal distance of around 600~km, with a SKR of $2.5 \cdot 10^{-2}$~bps (i.e. 2.2~kb per day (block)) at 600~km.
Going significantly beyond this limit would require switching to protocols featuring a more favorable dependency of the RKR as a function of the fiber length $l$, such as the recently proposed twin-field QKD ($\sim\exp(-l^{1/2})$)~\cite{Lucamarini2018}, or a quantum repeater~\cite{Sangouard2011}. However, these alternatives are of much greater technological complexity.


\section*{Acknowledgements}
We would like to acknowledge Jes\'us Mart\'inez-Mateo for providing the error correction code and Charles Ci Wen Lim for useful discussions.
We thank the Swiss NCCR QSIT.
D.R. and G.G. thank the EUs H2020 programme under the Marie Sk\l{}odowska-Curie project QCALL (GA 675662) for financial support.
This work was partly supported by the COST (European Cooperation in Science and Technology) Action MP1403 – Nanoscale Quantum Optics and by the Eurostars-2 joint programme (grant agreement: E11493 - QuPIC) with co-funding from the European Union Horizon 2020 research and innovation programme.

\bibliography{Bib_Stallion_LongDistance}
\end{document}